\documentclass[pdflatex,sn-mathphys-num]{sn-jnl}%
\usepackage{lineno}
\usepackage{graphicx}%
\usepackage{multirow}%
\usepackage{amsmath,amssymb,amsfonts}%
\usepackage{amsthm}%
\usepackage{mathrsfs}%
\usepackage[title]{appendix}%
\usepackage{xcolor}%
\usepackage{textcomp}%
\usepackage{manyfoot}%
\usepackage{booktabs}%
\usepackage{algorithm}%
\usepackage{algorithmicx}%
\usepackage{algpseudocode}%
\usepackage{listings}%
\usepackage{hyperref}
\usepackage{ulem}
\hypersetup{
    colorlinks=true,
    linkcolor=black,
    citecolor=blue,
    urlcolor=blue,
}
\usepackage[utf8]{inputenc}
\usepackage[T1]{fontenc}
\usepackage{mathptmx}
\usepackage{etoolbox}
\usepackage{hyperref}
\hypersetup{
    colorlinks=true,
    linkcolor=black,
    citecolor=blue,
    urlcolor=blue,
}

\begin{document}

\title[Article Title]{Harnessing Orbital Hall Effect in Spin-Orbit Torque MRAM}


\author*[1]{\fnm{Rahul} \sur{Gupta}}
\email{rahul.gupta.phy@outlook.com}

\author[2]{\fnm{Chloé} \sur{Bouard}}

\author[1]{\fnm{Fabian} \sur{Kammerbauer}}

\author[1]{\fnm{J. Omar} \sur{Ledesma-Martin}}

\author[1]{\fnm{Iryna } \sur{Kononenko}}

\author[2]{\fnm{Sylvain} \sur{Martin}}

\author[1,3]{\fnm{Gerhard} \sur{Jakob}}

\author[2]{\fnm{Marc} \sur{Drouard}}

\author*[1,3,4]{\fnm{Mathias} \sur{Kläui}}
\email{klaeui@uni-mainz.de}
\affil[1]{\orgdiv{Institute of Physics}, \orgname{Johannes Gutenberg University Mainz}, \orgaddress{\postcode{55099}, \city{Mainz},  \country{Germany}}}

\affil[2]{\orgname{Antaios}, \orgaddress{\postcode{38240}, \city{Meylan}, \country{France}}}

\affil[3]{\orgname{Graduate School of Excellence Materials Science in Mainz}, \orgaddress{ \postcode{55128}, \city{Mainz}, \country{Germany}}}

\affil[4]{\orgdiv{Department of Physics}, \orgname{Center for Quantum Spintronics, Norwegian University of Science and Technology}, \orgaddress{\postcode{7491}, \city{Trondheim}, \country{Norway}}}


\abstract{
Spin-Orbit Torque (SOT) Magnetic Random-Access Memory (MRAM) devices offer improved power efficiency, nonvolatility, and performance compared to static RAM, making them ideal, for instance, for cache memory applications. Efficient magnetization switching, long data retention, and high-density integration in SOT MRAM require ferromagnets (FM) with perpendicular magnetic anisotropy (PMA) combined with large torques enhanced by Orbital Hall Effect (OHE). We have engineered PMA [Co/Ni]$_3$ FM on selected OHE layers (Ru, Nb, Cr) and investigated the potential of theoretically predicted larger orbital Hall conductivity (OHC) to quantify the torque and switching current in OHE/[Co/Ni]$_3$ stacks. Our results demonstrate a $\sim$30\% enhancement in damping-like torque efficiency with a positive sign for the Ru OHE layer compared to a pure Pt, accompanied by a $\sim$20\% reduction in switching current for Ru compared to pure Pt across more than 250 devices, leading to more than a 60\% reduction in switching power. These findings validate the application of Ru in devices relevant to industrial contexts, supporting theoretical predictions regarding its superior OHC. This investigation highlights the potential of enhanced orbital torques to improve the performance of orbital-assisted SOT-MRAM, paving the way for next-generation memory technology.
}

\keywords{orbital Hall effect, switching current, perpendicular anisotropic magnet, SOT MRAM}

\maketitle
Data centers account for a substantial portion of global energy consumption, utilizing approximately 200 terawatt-hours annually, which represents currently 1\% of the world's energy use \cite{jones2018stop,andrae2015global}. To reduce the power consumption in computers, Spin-Orbit Torque (SOT) Magnetic Random-Access Memory (MRAM) is gaining attention as a potential replacement for static RAM due to its possible enhanced power efficiency, non-volatility, and superior performance, making it an attractive option for cache memory applications \cite{7008441,garello2018sot,kallinatha2024detailed}.

The operational principle of SOT MRAM involves current-induced magnetization switching as the write process \cite{RevModPhys.91.035004}. The efficiency of this switching, characterized by the product of switching power and switching time, is a critical determinant of the device's energy consumption. The switching power relies on the applied current required to switch the magnetization, which is referred to as the critical switching current (I$_c$). A lower critical current implies reduced power consumption in SOT devices. Moreover, the efficiency depends on three key factors: the conversion efficiency of charge to angular momentum within the SOT layer, the transfer of angular momentum at the interface, and the torque efficiency within the free layer of the magnetic tunnel junction. Another key aspect of SOT MRAM is its long data retention time, which depends on the thermal stability factor ($\Delta$). The industry typically requires $\Delta$ to be greater than 60, corresponding to 10 years of data retention \cite{7838492}.
Recent advancements have seen the use of $\beta$-W as an SOT layer in SOT MRAM for large-scale integration in ultrafast embedded memory applications \cite{garello2018sot}. Nonetheless, a challenge remains in the need for high switching current in $\beta$-W and its highly resistive nature ($\rho_{xx}$ = 200-300 $\mu \Omega$-cm \cite{pai2012spin}), which combined result in high power consumption. Consequently, high spin-orbit coupling (SOC) layers, such as Pt, are currently favored due to their significant spin Hall angle ($\theta_{SH}$ = 0.05-0.20 \cite{aradhya2016nanosecond,PhysRevLett.116.126601,cao2020prospect}), excellent conductivity ($\rho_{xx}$ = 15-50 $\mu \Omega$-cm \cite{PhysRevLett.116.126601,cao2020prospect}), and established intergration as a standard SOT layer. Recent proposals include various metallic alloys and multilayers as SOT layers to improve damping-like torque efficiency \cite{PtTialloySOT}. These alternatives focus on the transfer of spin angular momentum, necessitating materials with high SOC, thereby limiting the options to a few heavy elements in the periodic table, which are additionally often expensive and detrimental to the environment.

The Orbital Hall Effect (OHE) \cite{PhysRevLett.95.066601,go2021orbitronics,PhysRevLett.125.177201} and the Orbital Rashba Edelstein Effect (OREE) \cite{go2021orbitronics,PhysRevLett.128.067201,el2023observation} have been considered as promising mechanisms to enhance torques by tenfold \cite{PhysRevLett.125.177201}, without resorting to rare and expensive high SOC elements \cite{go2021orbitronics,PhysRevLett.125.177201,PhysRevLett.128.067201,PhysRevB.98.214405,PhysRevLett.102.016601,PhysRevLett.121.086602}, and are even argued as a fundamental mechanism of the Spin Hall Effect (SHE) \cite{PhysRevLett.102.016601,PhysRevLett.121.086602}. These phenomena leverage the generation, transfer, and conversion (from orbital to spin) of orbital angular momentum within SOT-based stacks. Theoretical studies suggest that the orbital Hall conductivity (OHC) can be significantly larger than spin Hall conductivities (SHC) in metals across the 3d, 4d, and 5d series \cite{salamiOHE,go2023first}. Recent experimental findings have corroborated the potential for enhanced torques due to the OHE and OREE \cite{PhysRevLett.125.177201,lee2021efficient,lee2021orbital,Petro3d4d5d,hayashi2023observation,LiuGiantOHE}. Previously, distinguishing between SHE and OHE has been achieved using in-plane magnetized ferromagnets (FM) such as Ni and FeCoB \cite{boselongrange}, which, however, are not viable for SOT-MRAM applications. It is worth noting that many of these studies have involved FM layers combined with OHE layers, where Rashba-type SOC effects at the interface \cite{baek2018spin,PhysRevLett.121.136805} and self-induced torques within the FM layer cannot be discounted \cite{wang2019anomalous,PhysRevB.99.220405}. Nonetheless, the presence of the OHE has been experimentally confirmed in bare non-FM materials with low SOC, such as Cr \cite{PMO_OHE_Cr} and Ti \cite{choi2023observation}, through magneto-optical Kerr detection. Moreover, it has been shown that the orbital angular momentum of an electron can propagate over longer distances than its spin counterpart \cite{hayashi2023observation,PhysRevResearch.5.023054,seifert2023time,xu2023orbitronics,boselongrange}.

Despite these strong experimental evidence and the notable torques achieved, harnessing the remarkable properties of OHE in industrial-scale SOT-MRAM devices poses significant challenges. Previous research has predominantly focused on oxide systems like Cu/CuO* \cite{an2016spin,PhysRevLett.125.177201,an2023electrical} and FMs with in-plane magnetization \cite{Petro3d4d5d,boselongrange,PhysRevB.106.144415,PhysRevResearch.5.023054}. In-plane magnetized systems do not lend themselves to high-density memories due to reduced thermal stability and stray fields, and the use of oxides results in significant power consumption due to the highly resistive nature of oxide layers. Oxides furthermore introduce complications in the fabrication processes due to the metal-oxide interface, thereby impeding the broad-scale integration of OHE-based technologies in industrial applications. Thus, reducing power consumption ($\propto I_c^2 \rho_{xx}$) involves reducing the switching current combined with excellent conductivity, which can be achieved by harnessing the OHE in SOT MRAM devices with all-metallic OHE stacks, along with sufficient perpendicular magnetic anisotropy (PMA) layers for high thermal stability.

In this article, we harness the OHE in SOT-MRAM devices at an industrially relevant scale to reduce the switching current when compared to traditional SHE-dominated SOT-MRAM devices (e.g., those employing Pt as an SOT layer). Our study incorporates fully metallic OHE layers, such as Ru, Nb, and Cr, alongside [Co/Ni]$_3$ as the perpendicularly magnetized FM layer in our SOT-MRAM devices. Through our industrial-scale fabrication process, we successfully probed more than 250 devices on full Si/SiO$_2$ wafers, revealing that Ru based SOT device provides a significantly larger torque as compared to Pt based SOT device. These experimental findings are consistent with the theoretically predicted higher OHC of Ru. Consequently, this leads to a reduction in the switching current observed in Ru/Pt/[Co/Ni]$_3$-based SOT MRAM devices compared to Pt/[Co/Ni]$_3$ devices,  especially in configurations with a thermal stability factor greater than 60. Our methodology involves the strategic integration of OHE layers into the writing process of SOT-MRAM devices. This integration includes ensuring that the stack exhibits sufficient PMA for high thermal stability, coupled with strong torques from the OHE and efficient conversion of orbital to spin, offering a promising avenue to reduce power consumption in data storage devices.\\
\\
\textbf{\Large  Perpendicular magnetization on OHE layers for high-density integration
}\\
Achieving PMA in combination with OHE layers is a key requirement for the architecture of highly dense and efficient SOT-MRAM devices. Therefore, multilayers of [Co(0.2)/Ni(0.6)]$_n$ were deposited onto selected OHE layers, such as Ru(2), Nb(2), and Cr(2), where \(n\) denotes the number of bilayer repetitions and the thickness values in nanometers are indicated in parentheses. A 1.5 nm thin Pt layer was inserted as an interlayer between the [Co/Ni]$_n$ (with n = 3 yielding optimal PMA) and the OHE layers to facilitate the induction of PMA, aligning with the requirements for SOT-MRAM devices, and serves as an efficient orbital-to-spin conversion layer due to its high SOC \cite{PhysRevLett.125.177201} (see section I, of the Supplementary Material (SM) and the methods section). This is crucial for the application of torque and the mechanisms of magnetization switching in these experiments.\\
\begin{figure*}[t!]
    \centering
    \includegraphics[width= 13cm]{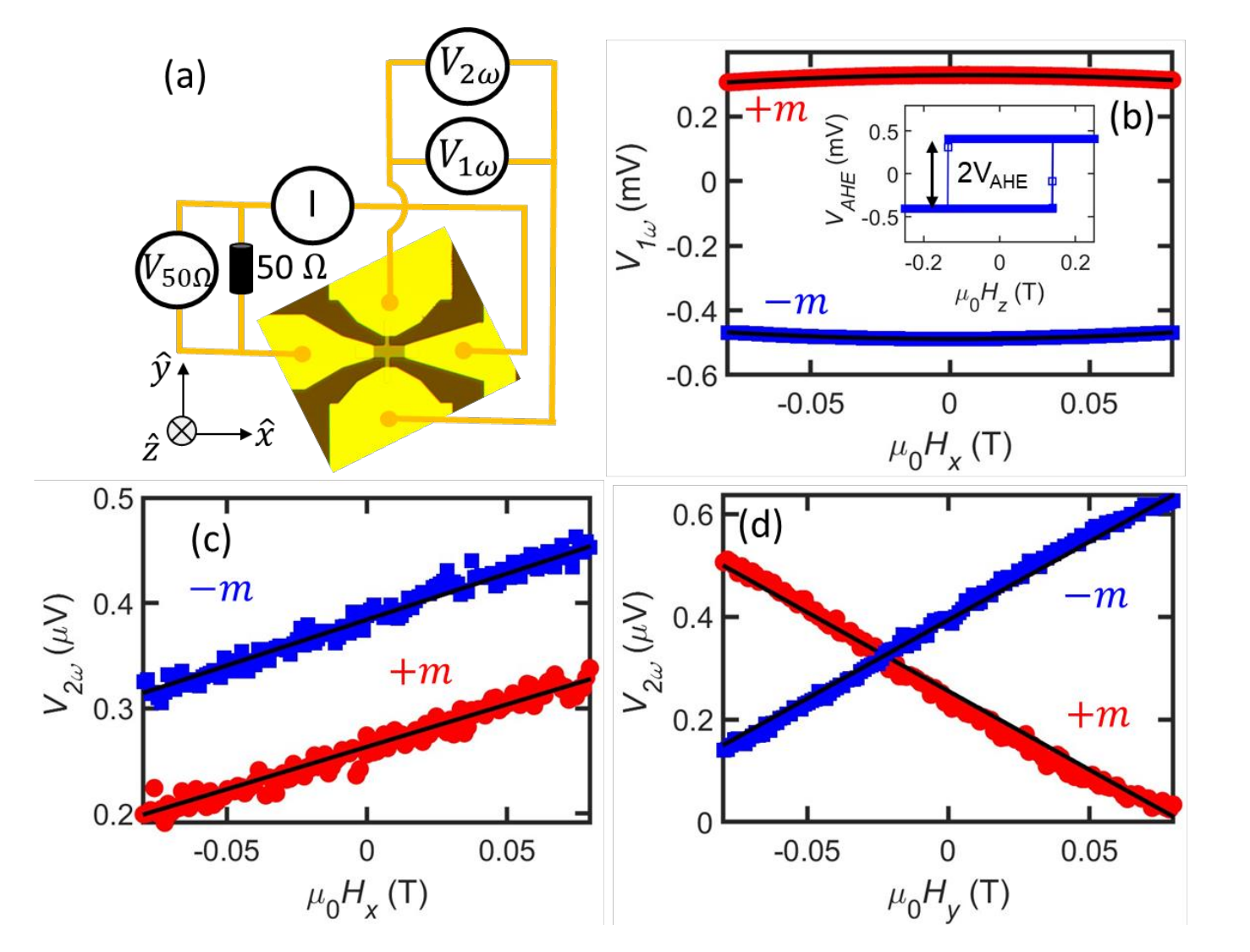}
    \caption{(a) Optical image of a device with the circuit diagram of the second harmonic Hall measurements. (b) First harmonic Hall voltage as a function of \(\mu_0H_x\). Inset shows the Anomalous Hall voltage as a function of \(\mu_0H_z\). (c) and (d) Second harmonic Hall voltage as a function of \(\mu_0H_x\) and \(\mu_0H_y\), respectively.
    }
\end{figure*}
\\
\textbf{\Large Enhanced torque due to orbital Hall effect}\\
To measure torques, we employ standard harmonic Hall measurements \cite{PhysRevB.89.144425} within a three-dimensional vector cryogenic setup, under the influence of applied magnetic fields at a temperature of 300K. Figure 1(a) shows the schematic diagram of the Hall bar device and its associated harmonic Hall measurement circuit. We apply a low-frequency (13.7 Hz) alternating electric current through the device and record both the first harmonic (V$_{1\omega}$) and second harmonic (V$_{2\omega}$) Hall voltages, systematically varying the magnetic field along the $\mu_0$H$_x$, $\mu_0$H$_y$, and $\mu_0$H$_z$ axes. Figure 1(b) displays V$_{1\omega}$ as a function of $\mu_0$H$_x$ for the Ru/Pt sample, demonstrating a change in the sign of V$_{1\omega}$ concurrent with the magnetization reversal. This behavior is consistent across $\mu_0$H$_y$ and extends to all tested samples. 
Before conducting these measurements, the samples were magnetically saturated out-of-plane by applying a field larger than their coercive field, establishing magnetization states denoted as +$m$ (up) and -$m$ (down). The inset of Fig. 1(b) depicts the anomalous Hall voltage (V$_{AHE}$) as a function of the out-of-plane magnetic field ($\mu_0$H$_z$), evidencing the PMA of [Co/Ni]$_3$ for all SOT layers; here data is specifically shown for the Ru/Pt OHE layer. 

Figures 1(c) and 1(d) present V$_{2\omega}$ as a function of $\mu_0$H$_x$ and $\mu_0$H$_y$ for magnetization states +$m$ and -$m$, respectively. These figures illustrate two distinct torque mechanisms in the [Co/Ni]$_3$ layers: 1) Damping-like (DL) torque, represented by $\tau_{DL} \propto m \times (\sigma \times m)$, where \(m\) is the unit vector of the Co/Ni magnetization and \(\sigma\) represents the angular momentum polarization due to SHE and/or OHE. 2) Field-like (FL) torque, denoted by $\tau_{FL} \propto m \times H_{eff}$, where $H_{eff}$ encompasses contributions from the interfacial spin-orbit field and the Oersted field due to conductive layers. The data in Fig. 1(c) is associated with DL torque, where V$_{2\omega}$ maintains its sign regardless of the magnetization direction, indicating DL torque characteristics. This pattern persists when the magnetic field and current are aligned (I$_{xx}$ $\parallel$ $\mu_0$H$_x$). In contrast, a change in the sign of V$_{2\omega}$ is observed when the field is perpendicular to the current (I$_{xx}$ $\bot$ $\mu_0$H$_y$), indicating the presence of FL torque as depicted in Fig. 1(d). However, when the planar Hall effect (PHE) becomes significant compared to the anomalous Hall effect (AHE), the efficiencies of DL ($\xi_{DL}^E$) and FL ($\xi_{FL}^E$) torques per unit electric field are defined as follows \cite{PhysRevB.89.144425},

\begin{equation}
    \xi_{DL(FL)}^E = \frac{2e}{\hbar} \frac{\mu_0 M_s t_{FM}}{E} \frac{H_{y(x)} \pm 2 \eta H_{x(y)}}{(1-4\eta^2)},
\end{equation}

where $2e$, $\hbar$, $\mu_0 M_s$, $E$, and $t_{FM}$ represent the charge of an electron, the reduced Planck constant, the saturation magnetization of the FM layer, the applied electric field, and the thickness of the FM layer, respectively. The parameter $\eta$ (= $\Delta R_{PHE}/\Delta R_{AHE}$) denotes the ratio of the resistance due to the PHE to that of the AHE. The measured fields, $H_{x(y)}$, are defined as:

\begin{equation}
    H_{x(y)} = -2 \left( {\frac{d V_{2\omega}}{dH_{x(y),app}}}\right) \left / \left({\frac{d^2 V_{1\omega}}{dH_{x(y),app}^2}}\right) \right .
\end{equation}

\begin{figure*}[b!]
    \centering
    \includegraphics[width= 13cm]{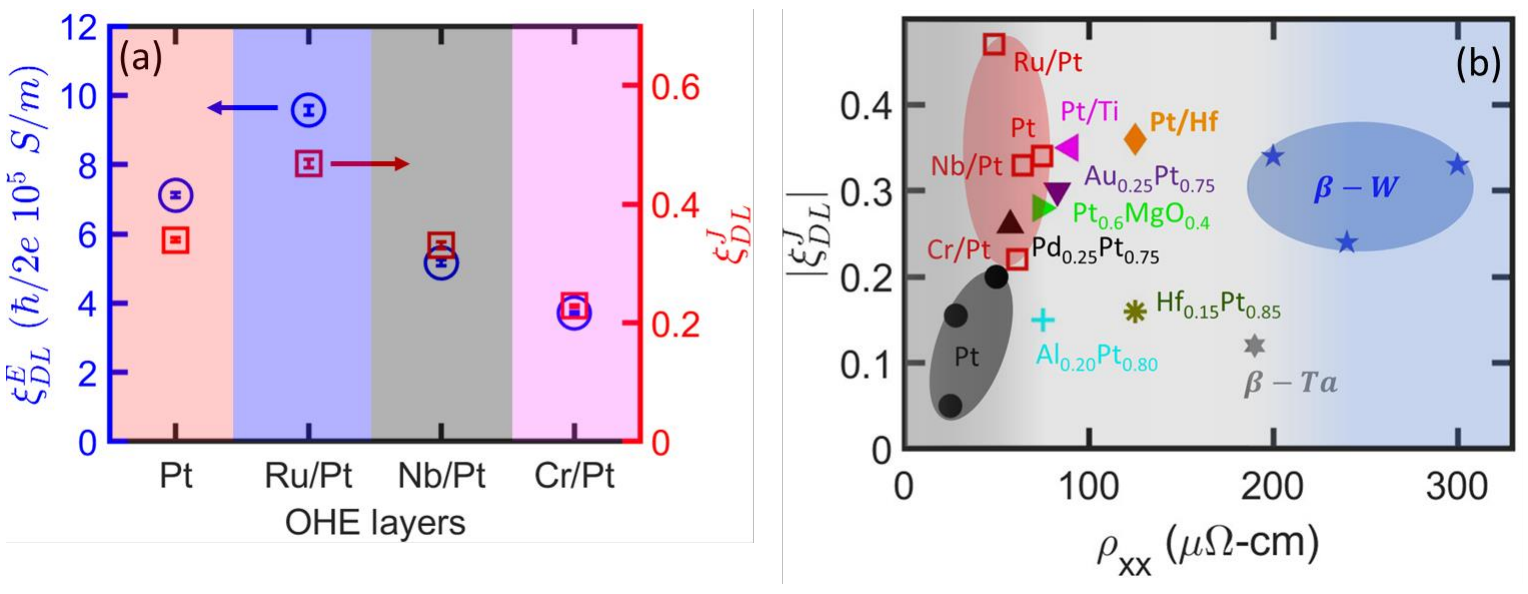}
    \caption{(a) Damping-like torque efficiency per unit electric field (left Y-axis) and per unit current density (right Y-axis) for various OHE-based heterostructures. (b) State-of-the-art comparison of damping-like torque efficiency per unit current density ($\xi_{DL(FL)}^J$) vs. longitudinal resistivity ($\rho_{xx}$) of NM layers. The values are taken for Pt \cite{PhysRevLett.116.126601,PhysRevLett.126.107204}, Pt/Ti \cite{PtTialloySOT}, Pt/Hf \cite{PhysRevApplied.11.061004}, PtAl alloy \cite{nguyen2016enhanced}, AuPt alloy \cite{PhysRevApplied.10.031001}, and $\beta$-W \cite{PhysRevApplied.9.011002,zhang2016critical}. Our data are represented by open square symbols in red.
    }
\end{figure*}

To evaluate the torque efficiencies, we initially measure $\mu_0 M_s$ of our samples using a Quantum Design Superconducting Quantum Interference Device (SQUID) (for details, see Section III, SM and the methods section). The measured $\mu_0 M_s$ values range from 0.92 (0.03) to 0.97 (0.04) Tesla across all samples, consistent with previously reported findings \cite{ito2022perpendicularly}. Subsequently, we measure the PHE as a function of the applied in-plane magnetic field, AHE as a function of out-of-plane applied field, and calculated the corresponding $\eta$ ratios for all samples, as described in the section II(S2,S3), SM. The resulting $\eta$ values for Pt, Ru/Pt, Nb/Pt, and Cr/Pt samples are found to be 0.72, 0.65, 0.86, and 1.04, respectively. Utilizing Eqs. (1-2), we deduce the torque efficiencies $\xi_{DL(FL)}^E$. These efficiencies, when expressed per unit current density for DL and FL torque components, are denoted as $\xi_{DL(FL)}^J = \xi_{DL(FL)}^E \rho_{NM}$, where $\rho_{NM}$ represents the resistivity of the SOT layer, consisting of Ta/Ru/Pt layers. The $\rho_{NM}$ values were determined employing a van der Pauw geometry, as described in Ref. \cite{oliveira2020simple} (for details, see the methods section).
Our findings indicate a predominance of a DL torque over a FL torque, with the FL torque being approximately 10\% of the DL torque across all samples (see section II, SM). 
As a key finding of our work, we discover that the torque efficiencies $\xi_{DL}^E$ and $\xi_{DL}^J$ exhibit a $\sim$30\% enhancement for the Ru/Pt OHE layer compared to the pure Pt layer, as illustrated in Fig. 2(a). Notably, this enhancement comes with a positive sign to the torque, indicating the dominant contribution of the OHE mechanism \cite{boselongrange}. This observation is in accordance with theoretical predictions suggesting a positive sign for the enhanced OHC in Ru, Nb, and Cr, as indicated by Salami et al. \cite{salamiOHE} and Go et al. \cite{go2023first}.

Figure 2(b) illustrates a comparison of $\xi_{DL}^J$ across various SOT layers, plotted as a function of their longitudinal resistivity ($\rho_{xx}$). Notably, the Ru/Pt SOT layer exhibits the highest torque efficiency compared to other investigated SOT materials, such as the Pt/Hf \cite{PhysRevApplied.11.061004}, the PtAl alloy \cite{nguyen2016enhanced}, Pt/Ti \cite{PtTialloySOT}, the AuPt alloy \cite{PhysRevApplied.10.031001}, and $\beta$-W \cite{PhysRevApplied.9.011002,zhang2016critical}. Recent studies have highlighted the potential of $\beta$-W as a material for SOT layer in SOT-MRAM applications, showcasing large-scale integration and a substantial value of $\xi_{DL}^J$ \cite{garello2018sot}. 
Nonetheless, its application is difficult due to the high power consumption resulting from its highly resistive nature ($\rho_{xx}$ = 200-300 $\mu \Omega$-cm). Thus, Ru/Pt stands out as a preferable SOT layer choice, outperforming pure Pt in torque efficiency and offering a much better conductivity than $\beta$-W, as illustrated in Fig. 2(b).\\
\begin{figure*}[t!]
    \centering
    \includegraphics[width=14cm]{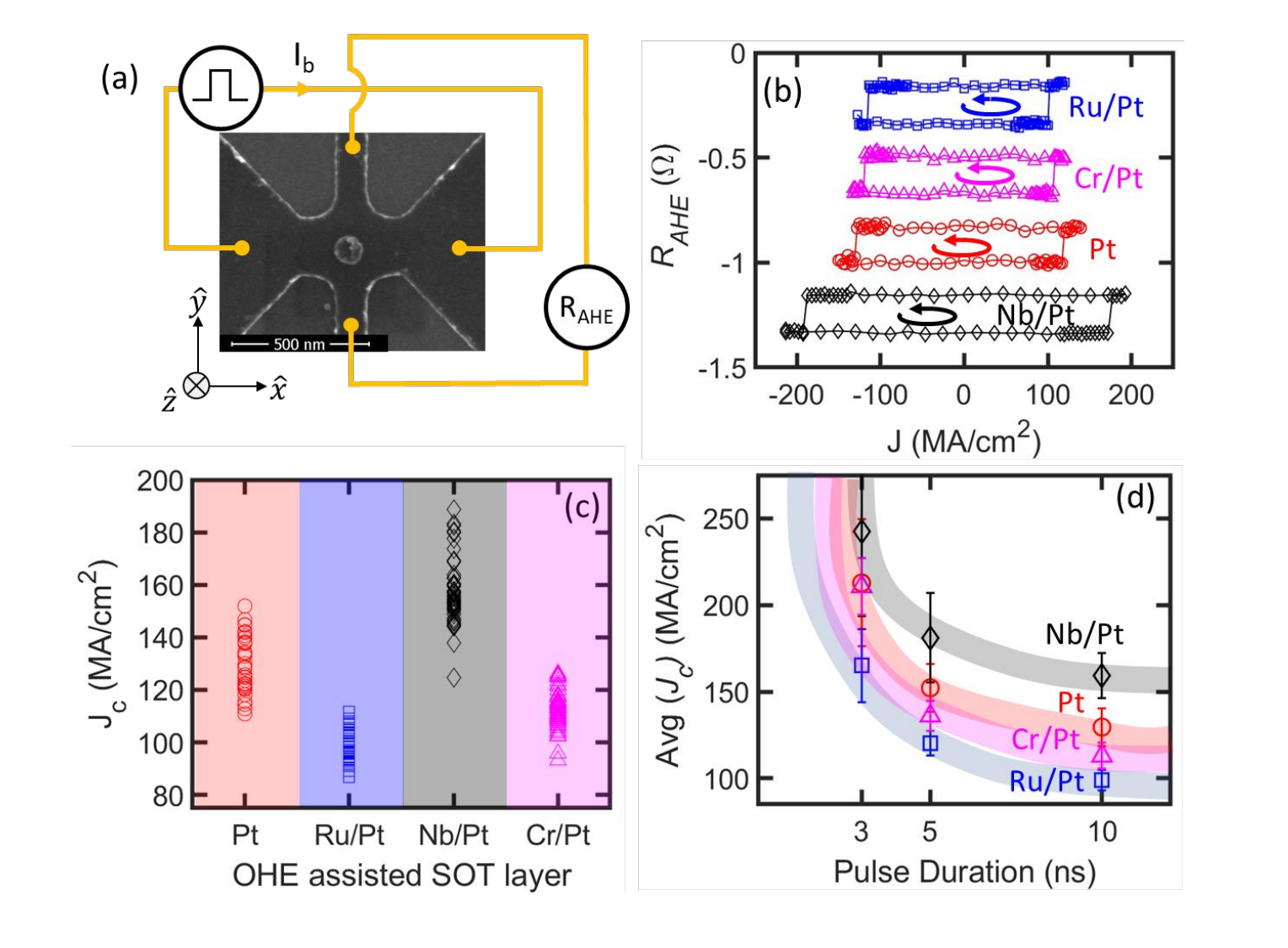}
    \caption{(a) Scanning electron microscope image with circuit diagram of the switching experiment. (b) Anomalous Hall resistance as a function of current density at a 10 ns pulse duration and \(\mu_0H_x\) = 50 mT. The y-axis has been rescaled for better visualization, and rounded arrows indicate the direction of switching polarity. (c) Critical current density for different OHE assisted SOT line at a 10 ns pulse duration and \(\mu_0H_x\) = 50 mT. (d) Average critical current density as a function of pulse duration at \(\mu_0H_x\) = 50 mT. Lines are guided by the eye.}
\end{figure*}
\\
\textbf{\Large Orbital-assisted current induced magnetization switching experiments: Analysis of >250 devices}\\
The primary figure of merit for a device is the critical current required for magnetization switching. Therefore, to examine the impact of enhanced torques on the switching current density, we fabricated SOT channels using bare Pt(3.5), Ru(2)/Pt(1.5), Nb(2)/Pt(1.5), and Cr(2)/Pt(1.5) as OHE layers, with dimensions of 200 $\times$ 400 nm$^2$. Note that the volume of the bare Pt layer is deliberately kept the same as the other SOT layers to ensure that the current density and current are considered equivalent in our work (refer to the methods section). Additionally, [Co/Ni]$_3$ was employed as a magnetic circular dot, with a diameter of 100 nm, for conducting switching experiments. The scanning electron microscopy image of a device, along with the circuit diagram, is depicted in Fig. 3(a). A rectangular pulse was applied in the $x$-direction, while the AHE voltage was measured along the $y$-direction, in the presence of an applied magnetic field $\mu_0 H_x$ of 50 mT. Figure 3(b) illustrates the AHE resistance (R$_{AHE}$) as a function of the applied current density ($J$) under a 50 mT magnetic field and with a 10 ns pulse duration. This graph reveals a notable decrease in the critical switching current density ($J_c$) with the use of Ru/Pt as an SOT layer in comparison to the bare Pt sample. Importantly, the switching polarity is consistent across all OHE layers (see Fig. 3b), corresponding to the positive sign observed in the torque efficiency measurements, indicating a dominant mechanism of OHE. Recognizing the significance of obtaining robust statistical data for precise analysis, we systematically probed more than 250 devices on an industrial scale on Si/SiO$_2$ wafers. This extensive effort has allowed us to attain a statistically significant understanding of $J_c$ in the samples. Subsequently, we graphically represent the $J_c$ for all OHE based devices in Fig. 3(c). Our large statistics underpins the reliability and precision of the findings reported in this study. In Fig. 3(d) we present the average $J_c$, computed across hundreds of devices, as a function of the applied current's pulse duration. The details for pulse durations of 3 ns and 5 ns are detailed in Section IV(S1) of the SM. In this analysis, the Ru/Pt OHE assisted SOT layer demonstrates a clear reduction of $\sim$20\% in the $J_c$ for devices scaled to industrial dimension, relative to those based purely on Pt SOT layer. It is important to note that $J_c$ can be influenced by factors such as the perpendicular anisotropic field ($H_k$), $\mu_0 M_s$, and $\xi_{DL}^J$ \cite{lee2013threshold}. In our investigation, we observed that the variation in these parameters across all the devices was confined to a 1-5\% range (see section III and IV(S2), SM), with the exception of $\xi_{DL}^J$, which is $\sim$30\%. This observation underscores that the observed $\sim$20\% decrease in average $J_c$ is predominantly attributable to the roughly $\sim$30\% improvement in DL torque efficiency. To gauge the key performance indicator of the power consumption, we ascertain the write switching power for our stacks, utilizing the framework described in Ref. \cite{PhysRevLett.126.107204}. Notably, the observed switching power in the Ru-based stack is reduced by 60\% compared to that measured for pure Pt and even a reduction of $\sim$220\% compared to $\beta$-W taking into account some published values of resistivity and switching current density \cite{garello2018sot}.
Finally and importantly, the thermal stability factors ($\Delta$) of our devices are found to be in the industrially acceptable range $\Delta$>60 (see section IV(S2), SM).\\
\\
{\textbf{\Large Conclusion}}\\
In conclusion, we have demonstrated the effective harnessing of the increased orbital Hall conductivity of Ru, Nb, and Cr layers in combination with a perpendicularly magnetized [Co/Ni]$_3$ ferromagnetic layer for Spin-Orbit Torque (SOT) Magnetic Random-Access Memory (MRAM) devices. This configuration enables efficient magnetization switching, suitable for high-density cache memory applications, for instance. Our findings reveal a significant $\sim$30\% enhancement in damping-like torque efficiency with positive sign for the Ru/Pt OHE layer compared to a pure Pt layer. Remarkably, this enhancement results in a $\sim$20\% reduction in switching current relative to a pure Pt layer across more than 250 devices, leading to a switching power reduction of more than 60\%. These results highlight the promising potential of leveraging the enhanced orbital Hall effect to propel the performance of next-generation of SOT MRAM devices for high-density packed cache memory applications.\\
\\
{\textbf{\Large Methods}}\\
{\textbf{\large Thin film deposition}}\\
The samples were prepared by DC and RF magnetron sputtering using a Singulus Rotaris sputtering tool, conducted at room temperature. The deposition was performed on the Si/SiO$_2$ substrates under a partial Argon pressure of 3.6-4.2$\times$10$^{-3}$ mbar, with the system maintaining a base pressure lower than 5$\times$10$^{-8}$ mbar. To ensure uniformity of the thin films, the sample was rotated at a speed of 60 rpm throughout the deposition process. The constructed sample stack comprised a Si/SiO$_2$ substrate followed by Ta(3) seed layer, selected OHE layers, a [Co(0.2)/Ni(0.6)]$_3$/Co(0.2) multilayer as a FM layer, MgO(2) as a barrier, CoFeB(0.3), Ta(1), and capped with Ru(5). The OHE layers included in the study were Pt(3.5), Ru(2)/Pt(1.5), Nb(2)/Pt(1.5), and Cr(2)/Pt(1.5), with the total thickness of the OHE layers maintained at 3.5 nm. The final stacks are as Substrate/Ta(3)/OHE(2)/Pt(1.5)/[Co(0.2)/Ni(0.6)]$_3$/Co(0.2)/MgO(2)/CoFeB(0.3)/Ta(1)/Ru(5) and Substrate/Ta(3)/Pt(3.5)/[Co(0.2)/Ni(0.6)]$_3$/Co(0.2)/MgO(2)/CoFeB(0.3)/Ta(1)/Ru(5) as a reference sample. \\
\\
{\textbf{\large Device fabrication}}\\
To fabricate the Hall bar for torque measurements, optical lithography was used to pattern the device into a Hall bar configuration, with dimensions of 30 $\times$ 45 $\mu m^2$ or 20 $\times$ 45 $\mu m^2$. This process was followed by a second step of optical lithography, after which sputtering and a lift-off process were conducted to establish the coplanar waveguide contact pads, as depicted in Figure 1(a). These contact pads consist of a 5 nm Cr layer capped with an 80 nm Au layer. Prior to the deposition of the contact pads, a mild argon ion etching was performed to ensure the establishment of transparent contacts.

For the current-induced magnetization switching experiments, hard mask crosses in Ti with a dimension of 200 $\times$ 400 nm$^2$ were initially patterned using ebeam lithography on wafers where the SOT layers had already been deposited. 
This step was followed by sputtering of Ti and lift-off. Subsequently, nanopillars (magnetic dots) with a diameter of 100 nm were created through electron lithography, Ti evaporation, and another lift-off process. The final step involved patterning and depositing contact pads made of Au (130 nm) on Ti (30 nm). The sample was then etched to reveal the devices, and photoresist was spin-coated onto the sample to shield the devices from oxidation.\\
\\
{\textbf{\large Resistivity measurements}}\\
The resistivity of the OHE assisted SOT layers was determined using the van der Pauw method. The SOT layers consisted of Ta(3)/Pt(3.5), Ta(3)/Ru(2)/Pt(1.5), Ta(3)/Nb(2)/Pt(1.5), and Ta(3)/Cr(2)/Pt(1.5). We applied a current using a Keithley 2400 source meter and measured resistances. In this procedure, four electrical contacts are placed on the sample. Current and voltage contacts are cycled through different switches.\\
\\
{\textbf{\large Magnetization measurements}}\\
A Quantum Design Superconducting Quantum Interference Device (SQUID) magnetometer was employed to measure the static magnetic moments of the samples by applying an in-plane magnetic field. To derive the magnetization, the measured moment values were normalized to the magnetic volume. This volume was determined by taking into account the film thickness—which was optimized by adjusting the deposition rate—and the precise area of the sample, which was measured using optical microscopy.\\
\\
{\textbf{\large Harmonic Hall experiments}}\\
For torque efficiency, we utilized the second-harmonic Hall measurement technique at room temperature in a three-dimensional vector cryostat. Standard wire bonding was used to connect the Hall bar device with the sample holder. The Hall bar is connected to a 50 $\Omega$ resistance in series to measure the input current during the measurement (see Fig. 1a). Before the harmonic measurements, the Hall bar device is pre-saturated along the $\hat{z}$-direction. We applied a sinusoidal voltage with constant amplitude, \(V_{in}(t) = V_0 \cos(2\pi ft)\), using a lock-in amplifier (HF2LI by Zurich Instruments) to the Hall bar device with a reference frequency (\(f\)) of 13.7 Hz. Two other lock-in amplifiers (signal recovery 7265 and 7225) were employed to simultaneously measure the in-phase first and out-of-phase second harmonic voltage, while sweeping the applied magnetic field in different orthogonal directions.\\
\\
{\textbf{\large Switching experiment}}\\
For the magnetization switching experiment, the devices were placed on an automated wafer prober outfitted with RF probes and a 3D magnet. A bias current, administered via a Keithley SMU 2450, was passed through the OHE-assisted SOT line of the device while subjected to a 50 mT in-plane magnetic field (\( \mu_0 H_x \)) parallel to the bias current. This setup allowed testing with various pulse durations—3, 5, or 10 ns—using an Active Technologies Pulse Rider AT PG-1047. The Anomalous Hall resistance was subsequently measured using a Keithley DMM 7510.

\backmatter
\bmhead{Data availability} The data that support the plots within this paper and other findings of this study are available from the corresponding author upon reasonable request.

\bmhead{Competing interests} The authors declare no competing interests.

\bmhead{Supplementary information} This article contains supplementary information.

\bmhead{Acknowledgements} R.G., F.K., O.S., G.J., and M.K. thank the DFG; Spin+X (A01, A11, B02) TRR 173-268565370 and Project No. 358671374; the Horizon 2020 Framework Programme of the European Commission under FETOpen Grant Agreement No. 863155 (s-Nebula); the European Research Council Grant Agreement No. 856538 (3D MAGiC); and the Research Council of Norway through its Centers of Excellence funding scheme, Project No. 262633 “QuSpin.”

\bmhead{Author contributions} 
R.G. played a primary role in performing device fabrication, harmonic Hall measurement, and data analysis for orbital torques under the supervision of M.K. F.K. deposited the thin films under the supervision of G.J. I.K. assisted F.K. in the film deposition. M.S. fabricated the devices for the switching experiment. C.B. conducted the switching experiment and performed preliminary analysis under the supervision of M.D. R.G. validated the analysis of switching data with the assistance of O.L. R.G. wrote the initial draft of the manuscript. All authors provided feedback on the manuscript and contributed to this work. M.K. has been the principal investigator of this project.

\bibliography{bibliography.bib}


\begin{thebibliography}{50}
\ifx \bisbn   \undefined \def \bisbn  #1{ISBN #1}\fi
\ifx \binits  \undefined \def \binits#1{#1}\fi
\ifx \bauthor  \undefined \def \bauthor#1{#1}\fi
\ifx \batitle  \undefined \def \batitle#1{#1}\fi
\ifx \bjtitle  \undefined \def \bjtitle#1{#1}\fi
\ifx \bvolume  \undefined \def \bvolume#1{\textbf{#1}}\fi
\ifx \byear  \undefined \def \byear#1{#1}\fi
\ifx \bissue  \undefined \def \bissue#1{#1}\fi
\ifx \bfpage  \undefined \def \bfpage#1{#1}\fi
\ifx \blpage  \undefined \def \blpage #1{#1}\fi
\ifx \burl  \undefined \def \burl#1{\textsf{#1}}\fi
\ifx \doiurl  \undefined \def \doiurl#1{\url{https://doi.org/#1}}\fi
\ifx \betal  \undefined \def \betal{\textit{et al.}}\fi
\ifx \binstitute  \undefined \def \binstitute#1{#1}\fi
\ifx \binstitutionaled  \undefined \def \binstitutionaled#1{#1}\fi
\ifx \bctitle  \undefined \def \bctitle#1{#1}\fi
\ifx \beditor  \undefined \def \beditor#1{#1}\fi
\ifx \bpublisher  \undefined \def \bpublisher#1{#1}\fi
\ifx \bbtitle  \undefined \def \bbtitle#1{#1}\fi
\ifx \bedition  \undefined \def \bedition#1{#1}\fi
\ifx \bseriesno  \undefined \def \bseriesno#1{#1}\fi
\ifx \blocation  \undefined \def \blocation#1{#1}\fi
\ifx \bsertitle  \undefined \def \bsertitle#1{#1}\fi
\ifx \bsnm \undefined \def \bsnm#1{#1}\fi
\ifx \bsuffix \undefined \def \bsuffix#1{#1}\fi
\ifx \bparticle \undefined \def \bparticle#1{#1}\fi
\ifx \barticle \undefined \def \barticle#1{#1}\fi
\bibcommenthead
\ifx \bconfdate \undefined \def \bconfdate #1{#1}\fi
\ifx \botherref \undefined \def \botherref #1{#1}\fi
\ifx \url \undefined \def \url#1{\textsf{#1}}\fi
\ifx \bchapter \undefined \def \bchapter#1{#1}\fi
\ifx \bbook \undefined \def \bbook#1{#1}\fi
\ifx \bcomment \undefined \def \bcomment#1{#1}\fi
\ifx \oauthor \undefined \def \oauthor#1{#1}\fi
\ifx \citeauthoryear \undefined \def \citeauthoryear#1{#1}\fi
\ifx \endbibitem  \undefined \def \endbibitem {}\fi
\ifx \bconflocation  \undefined \def \bconflocation#1{#1}\fi
\ifx \arxivurl  \undefined \def \arxivurl#1{\textsf{#1}}\fi
\csname PreBibitemsHook\endcsname

\bibitem[\protect\citeauthoryear{Jones et~al.}{2018}]{jones2018stop}
\begin{barticle}
\bauthor{\bsnm{Jones}, \binits{N.}}, \betal:
\batitle{How to stop data centres from gobbling up the world’s electricity}.
\bjtitle{Nature}
\bvolume{561}(\bissue{7722}),
\bfpage{163}--\blpage{166}
(\byear{2018})
\doiurl{10.1038/d41586-018-06610-y}
\end{barticle}
\endbibitem

\bibitem[\protect\citeauthoryear{Andrae and Edler}{2015}]{andrae2015global}
\begin{barticle}
\bauthor{\bsnm{Andrae}, \binits{A.S.}},
\bauthor{\bsnm{Edler}, \binits{T.}}:
\batitle{On global electricity usage of communication technology: trends to 2030}.
\bjtitle{Challenges}
\bvolume{6}(\bissue{1}),
\bfpage{117}--\blpage{157}
(\byear{2015})
\doiurl{10.3390/challe6010117}
\end{barticle}
\endbibitem

\bibitem[\protect\citeauthoryear{Oboril et~al.}{2015}]{7008441}
\begin{barticle}
\bauthor{\bsnm{Oboril}, \binits{F.}},
\bauthor{\bsnm{Bishnoi}, \binits{R.}},
\bauthor{\bsnm{Ebrahimi}, \binits{M.}},
\bauthor{\bsnm{Tahoori}, \binits{M.B.}}:
\batitle{Evaluation of hybrid memory technologies using \uppercase{SOT-MRAM} for on-chip cache hierarchy}.
\bjtitle{IEEE Trans. Comput.-Aided Des. Integr. Circuits Syst.}
\bvolume{34}(\bissue{3}),
\bfpage{367}--\blpage{380}
(\byear{2015})
\doiurl{10.1109/TCAD.2015.2391254}
\end{barticle}
\endbibitem

\bibitem[\protect\citeauthoryear{Garello et~al.}{2018}]{garello2018sot}
\begin{bchapter}
\bauthor{\bsnm{Garello}, \binits{K.}},
\bauthor{\bsnm{Yasin}, \binits{F.}},
\bauthor{\bsnm{Couet}, \binits{S.}},
\bauthor{\bsnm{Souriau}, \binits{L.}},
\bauthor{\bsnm{Swerts}, \binits{J.}},
\bauthor{\bsnm{Rao}, \binits{S.}},
\bauthor{\bsnm{Van~Beek}, \binits{S.}},
\bauthor{\bsnm{Kim}, \binits{W.}},
\bauthor{\bsnm{Liu}, \binits{E.}},
\bauthor{\bsnm{Kundu}, \binits{S.}}, \betal:
\bctitle{\uppercase{SOT-MRAM} 300mm integration for low power and ultrafast embedded memories}.
In: \bbtitle{2018 IEEE Symp. VLSI Circuits},
pp. \bfpage{81}--\blpage{82}
(\byear{2018}).
\doiurl{10.1109/VLSIC.2018.8502269} .
\bcomment{IEEE}.
\burl{https://ieeexplore.ieee.org/document/8502269}
\end{bchapter}
\endbibitem

\bibitem[\protect\citeauthoryear{Kallinatha et~al.}{2024}]{kallinatha2024detailed}
\begin{barticle}
\bauthor{\bsnm{Kallinatha}, \binits{H.}},
\bauthor{\bsnm{Rai}, \binits{S.}},
\bauthor{\bsnm{Talawar}, \binits{B.}}:
\batitle{A detailed study of \uppercase{SOT-MRAM} as an alternative to \uppercase{DRAM} primary memory in multi-core environment}.
\bjtitle{IEEE Access}
(\byear{2024})
\doiurl{10.1109/ACCESS.2024.3352151}
\end{barticle}
\endbibitem

\bibitem[\protect\citeauthoryear{Manchon et~al.}{2019}]{RevModPhys.91.035004}
\begin{barticle}
\bauthor{\bsnm{Manchon}, \binits{A.}},
\bauthor{\bsnm{\v{Z}elezn\'y}, \binits{J.}},
\bauthor{\bsnm{Miron}, \binits{I.M.}},
\bauthor{\bsnm{Jungwirth}, \binits{T.}},
\bauthor{\bsnm{Sinova}, \binits{J.}},
\bauthor{\bsnm{Thiaville}, \binits{A.}},
\bauthor{\bsnm{Garello}, \binits{K.}},
\bauthor{\bsnm{Gambardella}, \binits{P.}}:
\batitle{Current-induced spin-orbit torques in ferromagnetic and antiferromagnetic systems}.
\bjtitle{Rev. Mod. Phys.}
\bvolume{91},
\bfpage{035004}
(\byear{2019})
\doiurl{10.1103/RevModPhys.91.035004}
\end{barticle}
\endbibitem

\bibitem[\protect\citeauthoryear{Tillie et~al.}{2016}]{7838492}
\begin{bchapter}
\bauthor{\bsnm{Tillie}, \binits{L.}},
\bauthor{\bsnm{Nowak}, \binits{E.}},
\bauthor{\bsnm{Sousa}, \binits{R.C.}},
\bauthor{\bsnm{Cyrille}, \binits{M.-C.}},
\bauthor{\bsnm{Delaet}, \binits{B.}},
\bauthor{\bsnm{Magis}, \binits{T.}},
\bauthor{\bsnm{Persico}, \binits{A.}},
\bauthor{\bsnm{Langer}, \binits{J.}},
\bauthor{\bsnm{Ocker}, \binits{B.}},
\bauthor{\bsnm{Prejbeanu}, \binits{I.-L.}},
\bauthor{\bsnm{Perniola}, \binits{L.}}:
\bctitle{Data retention extraction methodology for perpendicular \uppercase{STT-MRAM}}.
In: \bbtitle{2016 IEEE International Electron Devices Meeting (IEDM)},
pp. \bfpage{27}--\blpage{312734}
(\byear{2016}).
\doiurl{10.1109/IEDM.2016.7838492}
\end{bchapter}
\endbibitem

\bibitem[\protect\citeauthoryear{Pai et~al.}{2012}]{pai2012spin}
\begin{botherref}
\oauthor{\bsnm{Pai}, \binits{C.-F.}},
\oauthor{\bsnm{Liu}, \binits{L.}},
\oauthor{\bsnm{Li}, \binits{Y.}},
\oauthor{\bsnm{Tseng}, \binits{H.}},
\oauthor{\bsnm{Ralph}, \binits{D.}},
\oauthor{\bsnm{Buhrman}, \binits{R.}}:
Spin transfer torque devices utilizing the giant spin \uppercase{H}all effect of tungsten.
Appl. Phys. Lett.
\textbf{101}(12)
(2012)
\doiurl{10.1063/1.4753947}
\end{botherref}
\endbibitem

\bibitem[\protect\citeauthoryear{Aradhya et~al.}{2016}]{aradhya2016nanosecond}
\begin{barticle}
\bauthor{\bsnm{Aradhya}, \binits{S.V.}},
\bauthor{\bsnm{Rowlands}, \binits{G.E.}},
\bauthor{\bsnm{Oh}, \binits{J.}},
\bauthor{\bsnm{Ralph}, \binits{D.C.}},
\bauthor{\bsnm{Buhrman}, \binits{R.A.}}:
\batitle{Nanosecond-timescale low energy switching of in-plane magnetic tunnel junctions through dynamic oersted-field-assisted spin \uppercase{H}all effect}.
\bjtitle{Nano Lett.}
\bvolume{16}(\bissue{10}),
\bfpage{5987}--\blpage{5992}
(\byear{2016})
\doiurl{10.1021/acs.nanolett.6b01443}
\end{barticle}
\endbibitem

\bibitem[\protect\citeauthoryear{Nguyen et~al.}{2016}]{PhysRevLett.116.126601}
\begin{barticle}
\bauthor{\bsnm{Nguyen}, \binits{M.-H.}},
\bauthor{\bsnm{Ralph}, \binits{D.C.}},
\bauthor{\bsnm{Buhrman}, \binits{R.A.}}:
\batitle{Spin torque study of the spin \uppercase{H}all conductivity and spin diffusion length in platinum thin films with varying resistivity}.
\bjtitle{Phys. Rev. Lett.}
\bvolume{116},
\bfpage{126601}
(\byear{2016})
\doiurl{10.1103/PhysRevLett.116.126601}
\end{barticle}
\endbibitem

\bibitem[\protect\citeauthoryear{Cao et~al.}{2020}]{cao2020prospect}
\begin{botherref}
\oauthor{\bsnm{Cao}, \binits{Y.}},
\oauthor{\bsnm{Xing}, \binits{G.}},
\oauthor{\bsnm{Lin}, \binits{H.}},
\oauthor{\bsnm{Zhang}, \binits{N.}},
\oauthor{\bsnm{Zheng}, \binits{H.}},
\oauthor{\bsnm{Wang}, \binits{K.}}:
Prospect of spin-orbitronic devices and their applications.
iScience
\textbf{23}(10)
(2020)
\doiurl{10.1016/j.isci.2020.101614}
\end{botherref}
\endbibitem

\bibitem[\protect\citeauthoryear{Zhu and Buhrman}{2019}]{PtTialloySOT}
\begin{barticle}
\bauthor{\bsnm{Zhu}, \binits{L.}},
\bauthor{\bsnm{Buhrman}, \binits{R.A.}}:
\batitle{Maximizing spin-orbit-torque efficiency of $\mathrm{Pt}/\mathrm{Ti}$ multilayers: \uppercase{T}rade-off between intrinsic spin \uppercase{H}all conductivity and carrier lifetime}.
\bjtitle{Phys. Rev. Appl.}
\bvolume{12},
\bfpage{051002}
(\byear{2019})
\doiurl{10.1103/PhysRevApplied.12.051002}
\end{barticle}
\endbibitem

\bibitem[\protect\citeauthoryear{Bernevig et~al.}{2005}]{PhysRevLett.95.066601}
\begin{barticle}
\bauthor{\bsnm{Bernevig}, \binits{B.A.}},
\bauthor{\bsnm{Hughes}, \binits{T.L.}},
\bauthor{\bsnm{Zhang}, \binits{S.-C.}}:
\batitle{Orbitronics: \uppercase{T}he intrinsic orbital current in $p$-doped silicon}.
\bjtitle{Phys. Rev. Lett.}
\bvolume{95},
\bfpage{066601}
(\byear{2005})
\doiurl{10.1103/PhysRevLett.95.066601}
\end{barticle}
\endbibitem

\bibitem[\protect\citeauthoryear{Go et~al.}{2021}]{go2021orbitronics}
\begin{barticle}
\bauthor{\bsnm{Go}, \binits{D.}},
\bauthor{\bsnm{Jo}, \binits{D.}},
\bauthor{\bsnm{Lee}, \binits{H.-W.}},
\bauthor{\bsnm{Kl{\"a}ui}, \binits{M.}},
\bauthor{\bsnm{Mokrousov}, \binits{Y.}}:
\batitle{Orbitronics: \uppercase{O}rbital currents in solids}.
\bjtitle{Europhys. Lett.}
\bvolume{135}(\bissue{3}),
\bfpage{37001}
(\byear{2021})
\doiurl{10.1209/0295-5075/ac2653}
\end{barticle}
\endbibitem

\bibitem[\protect\citeauthoryear{Ding et~al.}{2020}]{PhysRevLett.125.177201}
\begin{barticle}
\bauthor{\bsnm{Ding}, \binits{S.}},
\bauthor{\bsnm{Ross}, \binits{A.}},
\bauthor{\bsnm{Go}, \binits{D.}},
\bauthor{\bsnm{Baldrati}, \binits{L.}},
\bauthor{\bsnm{Ren}, \binits{Z.}},
\bauthor{\bsnm{Freimuth}, \binits{F.}},
\bauthor{\bsnm{Becker}, \binits{S.}},
\bauthor{\bsnm{Kammerbauer}, \binits{F.}},
\bauthor{\bsnm{Yang}, \binits{J.}},
\bauthor{\bsnm{Jakob}, \binits{G.}},
\bauthor{\bsnm{Mokrousov}, \binits{Y.}},
\bauthor{\bsnm{Kl\"aui}, \binits{M.}}:
\batitle{Harnessing orbital-to-spin conversion of interfacial orbital currents for efficient spin-orbit torques}.
\bjtitle{Phys. Rev. Lett.}
\bvolume{125},
\bfpage{177201}
(\byear{2020})
\doiurl{10.1103/PhysRevLett.125.177201}
\end{barticle}
\endbibitem

\bibitem[\protect\citeauthoryear{Ding et~al.}{2022}]{PhysRevLett.128.067201}
\begin{barticle}
\bauthor{\bsnm{Ding}, \binits{S.}},
\bauthor{\bsnm{Liang}, \binits{Z.}},
\bauthor{\bsnm{Go}, \binits{D.}},
\bauthor{\bsnm{Yun}, \binits{C.}},
\bauthor{\bsnm{Xue}, \binits{M.}},
\bauthor{\bsnm{Liu}, \binits{Z.}},
\bauthor{\bsnm{Becker}, \binits{S.}},
\bauthor{\bsnm{Yang}, \binits{W.}},
\bauthor{\bsnm{Du}, \binits{H.}},
\bauthor{\bsnm{Wang}, \binits{C.}},
\bauthor{\bsnm{Yang}, \binits{Y.}},
\bauthor{\bsnm{Jakob}, \binits{G.}},
\bauthor{\bsnm{Kl\"aui}, \binits{M.}},
\bauthor{\bsnm{Mokrousov}, \binits{Y.}},
\bauthor{\bsnm{Yang}, \binits{J.}}:
\batitle{Observation of the orbital \uppercase{R}ashba-\uppercase{E}delstein magnetoresistance}.
\bjtitle{Phys. Rev. Lett.}
\bvolume{128},
\bfpage{067201}
(\byear{2022})
\doiurl{10.1103/PhysRevLett.128.067201}
\end{barticle}
\endbibitem

\bibitem[\protect\citeauthoryear{El~Hamdi et~al.}{2023}]{el2023observation}
\begin{barticle}
\bauthor{\bsnm{El~Hamdi}, \binits{A.}},
\bauthor{\bsnm{Chauleau}, \binits{J.-Y.}},
\bauthor{\bsnm{Boselli}, \binits{M.}},
\bauthor{\bsnm{Thibault}, \binits{C.}},
\bauthor{\bsnm{Gorini}, \binits{C.}},
\bauthor{\bsnm{Smogunov}, \binits{A.}},
\bauthor{\bsnm{Barreteau}, \binits{C.}},
\bauthor{\bsnm{Gariglio}, \binits{S.}},
\bauthor{\bsnm{Triscone}, \binits{J.-M.}},
\bauthor{\bsnm{Viret}, \binits{M.}}:
\batitle{Observation of the orbital inverse \uppercase{R}ashba--\uppercase{E}delstein effect}.
\bjtitle{Nat. Phys.}
\bvolume{19}(\bissue{12}),
\bfpage{1855}--\blpage{1860}
(\byear{2023})
\doiurl{10.1038/s41567-023-02121-4}
\end{barticle}
\endbibitem

\bibitem[\protect\citeauthoryear{Jo et~al.}{2018}]{PhysRevB.98.214405}
\begin{barticle}
\bauthor{\bsnm{Jo}, \binits{D.}},
\bauthor{\bsnm{Go}, \binits{D.}},
\bauthor{\bsnm{Lee}, \binits{H.-W.}}:
\batitle{Gigantic intrinsic orbital \uppercase{H}all effects in weakly spin-orbit coupled metals}.
\bjtitle{Phys. Rev. B}
\bvolume{98},
\bfpage{214405}
(\byear{2018})
\doiurl{10.1103/PhysRevB.98.214405}
\end{barticle}
\endbibitem

\bibitem[\protect\citeauthoryear{Kontani et~al.}{2009}]{PhysRevLett.102.016601}
\begin{barticle}
\bauthor{\bsnm{Kontani}, \binits{H.}},
\bauthor{\bsnm{Tanaka}, \binits{T.}},
\bauthor{\bsnm{Hirashima}, \binits{D.S.}},
\bauthor{\bsnm{Yamada}, \binits{K.}},
\bauthor{\bsnm{Inoue}, \binits{J.}}:
\batitle{Giant orbital \uppercase{H}all effect in transition metals: \uppercase{O}rigin of large spin and anomalous \uppercase{H}all effects}.
\bjtitle{Phys. Rev. Lett.}
\bvolume{102},
\bfpage{016601}
(\byear{2009})
\doiurl{10.1103/PhysRevLett.102.016601}
\end{barticle}
\endbibitem

\bibitem[\protect\citeauthoryear{Go et~al.}{2018}]{PhysRevLett.121.086602}
\begin{barticle}
\bauthor{\bsnm{Go}, \binits{D.}},
\bauthor{\bsnm{Jo}, \binits{D.}},
\bauthor{\bsnm{Kim}, \binits{C.}},
\bauthor{\bsnm{Lee}, \binits{H.-W.}}:
\batitle{Intrinsic spin and orbital \uppercase{H}all effects from orbital texture}.
\bjtitle{Phys. Rev. Lett.}
\bvolume{121},
\bfpage{086602}
(\byear{2018})
\doiurl{10.1103/PhysRevLett.121.086602}
\end{barticle}
\endbibitem

\bibitem[\protect\citeauthoryear{Salemi and Oppeneer}{2022}]{salamiOHE}
\begin{barticle}
\bauthor{\bsnm{Salemi}, \binits{L.}},
\bauthor{\bsnm{Oppeneer}, \binits{P.M.}}:
\batitle{First-principles theory of intrinsic spin and orbital \uppercase{H}all and \uppercase{N}ernst effects in metallic monoatomic crystals}.
\bjtitle{Phys. Rev. Mater.}
\bvolume{6},
\bfpage{095001}
(\byear{2022})
\doiurl{10.1103/PhysRevMaterials.6.095001}
\end{barticle}
\endbibitem

\bibitem[\protect\citeauthoryear{Go et~al.}{2023}]{go2023first}
\begin{barticle}
\bauthor{\bsnm{Go}, \binits{D.}},
\bauthor{\bsnm{Lee}, \binits{H.-W.}},
\bauthor{\bsnm{Oppeneer}, \binits{P.M.}},
\bauthor{\bsnm{Bl{\"u}gel}, \binits{S.}},
\bauthor{\bsnm{Mokrousov}, \binits{Y.}}:
\batitle{First-principles calculation of orbital \uppercase{H}all effect by wannier interpolation: Role of orbital dependence of the anomalous position}.
\bjtitle{arXiv preprint arXiv:2309.13996}
(\byear{2023})
\doiurl{10.48550/arXiv.2309.13996}
\end{barticle}
\endbibitem

\bibitem[\protect\citeauthoryear{Lee et~al.}{2021a}]{lee2021efficient}
\begin{barticle}
\bauthor{\bsnm{Lee}, \binits{S.}},
\bauthor{\bsnm{Kang}, \binits{M.-G.}},
\bauthor{\bsnm{Go}, \binits{D.}},
\bauthor{\bsnm{Kim}, \binits{D.}},
\bauthor{\bsnm{Kang}, \binits{J.-H.}},
\bauthor{\bsnm{Lee}, \binits{T.}},
\bauthor{\bsnm{Lee}, \binits{G.-H.}},
\bauthor{\bsnm{Kang}, \binits{J.}},
\bauthor{\bsnm{Lee}, \binits{N.J.}},
\bauthor{\bsnm{Mokrousov}, \binits{Y.}}, \betal:
\batitle{Efficient conversion of orbital \uppercase{H}all current to spin current for spin-orbit torque switching}.
\bjtitle{Commun. Phys.}
\bvolume{4}(\bissue{1}),
\bfpage{234}
(\byear{2021})
\doiurl{10.1038/s42005-021-00737-7}
\end{barticle}
\endbibitem

\bibitem[\protect\citeauthoryear{Lee et~al.}{2021b}]{lee2021orbital}
\begin{barticle}
\bauthor{\bsnm{Lee}, \binits{D.}},
\bauthor{\bsnm{Go}, \binits{D.}},
\bauthor{\bsnm{Park}, \binits{H.-J.}},
\bauthor{\bsnm{Jeong}, \binits{W.}},
\bauthor{\bsnm{Ko}, \binits{H.-W.}},
\bauthor{\bsnm{Yun}, \binits{D.}},
\bauthor{\bsnm{Jo}, \binits{D.}},
\bauthor{\bsnm{Lee}, \binits{S.}},
\bauthor{\bsnm{Go}, \binits{G.}},
\bauthor{\bsnm{Oh}, \binits{J.H.}}, \betal:
\batitle{Orbital torque in magnetic bilayers}.
\bjtitle{Nat. Commun.}
\bvolume{12}(\bissue{1}),
\bfpage{6710}
(\byear{2021})
\doiurl{10.1038/s41467-021-26650-9}
\end{barticle}
\endbibitem

\bibitem[\protect\citeauthoryear{Sala and Gambardella}{2022}]{Petro3d4d5d}
\begin{barticle}
\bauthor{\bsnm{Sala}, \binits{G.}},
\bauthor{\bsnm{Gambardella}, \binits{P.}}:
\batitle{Giant orbital \uppercase{H}all effect and orbital-to-spin conversion in $3d$, $5d$, and $4f$ metallic heterostructures}.
\bjtitle{Phys. Rev. Res.}
\bvolume{4},
\bfpage{033037}
(\byear{2022})
\doiurl{10.1103/PhysRevResearch.4.033037}
\end{barticle}
\endbibitem

\bibitem[\protect\citeauthoryear{Hayashi et~al.}{2023}]{hayashi2023observation}
\begin{barticle}
\bauthor{\bsnm{Hayashi}, \binits{H.}},
\bauthor{\bsnm{Jo}, \binits{D.}},
\bauthor{\bsnm{Go}, \binits{D.}},
\bauthor{\bsnm{Gao}, \binits{T.}},
\bauthor{\bsnm{Haku}, \binits{S.}},
\bauthor{\bsnm{Mokrousov}, \binits{Y.}},
\bauthor{\bsnm{Lee}, \binits{H.-W.}},
\bauthor{\bsnm{Ando}, \binits{K.}}:
\batitle{Observation of long-range orbital transport and giant orbital torque}.
\bjtitle{Commun. Phys.}
\bvolume{6}(\bissue{1}),
\bfpage{32}
(\byear{2023})
\doiurl{10.1038/s42005-023-01139-7}
\end{barticle}
\endbibitem

\bibitem[\protect\citeauthoryear{Liu et~al.}{2023}]{LiuGiantOHE}
\begin{barticle}
\bauthor{\bsnm{Liu}, \binits{F.}},
\bauthor{\bsnm{Liang}, \binits{B.}},
\bauthor{\bsnm{Xu}, \binits{J.}},
\bauthor{\bsnm{Jia}, \binits{C.}},
\bauthor{\bsnm{Jiang}, \binits{C.}}:
\batitle{Giant efficiency of long-range orbital torque in \uppercase{C}o/\uppercase{N}b bilayers}.
\bjtitle{Phys. Rev. B}
\bvolume{107},
\bfpage{054404}
(\byear{2023})
\doiurl{10.1103/PhysRevB.107.054404}
\end{barticle}
\endbibitem

\bibitem[\protect\citeauthoryear{Bose et~al.}{2023}]{boselongrange}
\begin{barticle}
\bauthor{\bsnm{Bose}, \binits{A.}},
\bauthor{\bsnm{Kammerbauer}, \binits{F.}},
\bauthor{\bsnm{Gupta}, \binits{R.}},
\bauthor{\bsnm{Go}, \binits{D.}},
\bauthor{\bsnm{Mokrousov}, \binits{Y.}},
\bauthor{\bsnm{Jakob}, \binits{G.}},
\bauthor{\bsnm{Kl\"aui}, \binits{M.}}:
\batitle{Detection of long-range orbital-\uppercase{H}all torques}.
\bjtitle{Phys. Rev. B}
\bvolume{107},
\bfpage{134423}
(\byear{2023})
\doiurl{10.1103/PhysRevB.107.134423}
\end{barticle}
\endbibitem

\bibitem[\protect\citeauthoryear{Baek et~al.}{2018}]{baek2018spin}
\begin{barticle}
\bauthor{\bsnm{Baek}, \binits{S.-h.C.}},
\bauthor{\bsnm{Amin}, \binits{V.P.}},
\bauthor{\bsnm{Oh}, \binits{Y.-W.}},
\bauthor{\bsnm{Go}, \binits{G.}},
\bauthor{\bsnm{Lee}, \binits{S.-J.}},
\bauthor{\bsnm{Lee}, \binits{G.-H.}},
\bauthor{\bsnm{Kim}, \binits{K.-J.}},
\bauthor{\bsnm{Stiles}, \binits{M.D.}},
\bauthor{\bsnm{Park}, \binits{B.-G.}},
\bauthor{\bsnm{Lee}, \binits{K.-J.}}:
\batitle{Spin currents and spin-orbit torques in ferromagnetic trilayers}.
\bjtitle{Nat. Mater.}
\bvolume{17}(\bissue{6}),
\bfpage{509}--\blpage{513}
(\byear{2018})
\doiurl{10.1038/s41563-018-0041-5}
\end{barticle}
\endbibitem

\bibitem[\protect\citeauthoryear{Amin et~al.}{2018}]{PhysRevLett.121.136805}
\begin{barticle}
\bauthor{\bsnm{Amin}, \binits{V.P.}},
\bauthor{\bsnm{Zemen}, \binits{J.}},
\bauthor{\bsnm{Stiles}, \binits{M.D.}}:
\batitle{Interface-generated spin currents}.
\bjtitle{Phys. Rev. Lett.}
\bvolume{121},
\bfpage{136805}
(\byear{2018})
\doiurl{10.1103/PhysRevLett.121.136805}
\end{barticle}
\endbibitem

\bibitem[\protect\citeauthoryear{Wang et~al.}{2019}]{wang2019anomalous}
\begin{barticle}
\bauthor{\bsnm{Wang}, \binits{W.}},
\bauthor{\bsnm{Wang}, \binits{T.}},
\bauthor{\bsnm{Amin}, \binits{V.P.}},
\bauthor{\bsnm{Wang}, \binits{Y.}},
\bauthor{\bsnm{Radhakrishnan}, \binits{A.}},
\bauthor{\bsnm{Davidson}, \binits{A.}},
\bauthor{\bsnm{Allen}, \binits{S.R.}},
\bauthor{\bsnm{Silva}, \binits{T.J.}},
\bauthor{\bsnm{Ohldag}, \binits{H.}},
\bauthor{\bsnm{Balzar}, \binits{D.}}, \betal:
\batitle{Anomalous spin-orbit torques in magnetic single-layer films}.
\bjtitle{Nat. Nanotechnol.}
\bvolume{14}(\bissue{9}),
\bfpage{819}--\blpage{824}
(\byear{2019})
\doiurl{10.1038/s41565-019-0504-0}
\end{barticle}
\endbibitem

\bibitem[\protect\citeauthoryear{Amin et~al.}{2019}]{PhysRevB.99.220405}
\begin{barticle}
\bauthor{\bsnm{Amin}, \binits{V.P.}},
\bauthor{\bsnm{Li}, \binits{J.}},
\bauthor{\bsnm{Stiles}, \binits{M.D.}},
\bauthor{\bsnm{Haney}, \binits{P.M.}}:
\batitle{Intrinsic spin currents in ferromagnets}.
\bjtitle{Phys. Rev. B}
\bvolume{99},
\bfpage{220405}
(\byear{2019})
\doiurl{10.1103/PhysRevB.99.220405}
\end{barticle}
\endbibitem

\bibitem[\protect\citeauthoryear{Lyalin et~al.}{2023}]{PMO_OHE_Cr}
\begin{barticle}
\bauthor{\bsnm{Lyalin}, \binits{I.}},
\bauthor{\bsnm{Alikhah}, \binits{S.}},
\bauthor{\bsnm{Berritta}, \binits{M.}},
\bauthor{\bsnm{Oppeneer}, \binits{P.M.}},
\bauthor{\bsnm{Kawakami}, \binits{R.K.}}:
\batitle{Magneto-optical detection of the orbital \uppercase{H}all effect in chromium}.
\bjtitle{Phys. Rev. Lett.}
\bvolume{131},
\bfpage{156702}
(\byear{2023})
\doiurl{10.1103/PhysRevLett.131.156702}
\end{barticle}
\endbibitem

\bibitem[\protect\citeauthoryear{Choi et~al.}{2023}]{choi2023observation}
\begin{barticle}
\bauthor{\bsnm{Choi}, \binits{Y.-G.}},
\bauthor{\bsnm{Jo}, \binits{D.}},
\bauthor{\bsnm{Ko}, \binits{K.-H.}},
\bauthor{\bsnm{Go}, \binits{D.}},
\bauthor{\bsnm{Kim}, \binits{K.-H.}},
\bauthor{\bsnm{Park}, \binits{H.G.}},
\bauthor{\bsnm{Kim}, \binits{C.}},
\bauthor{\bsnm{Min}, \binits{B.-C.}},
\bauthor{\bsnm{Choi}, \binits{G.-M.}},
\bauthor{\bsnm{Lee}, \binits{H.-W.}}:
\batitle{Observation of the orbital \uppercase{H}all effect in a light metal \uppercase{T}i}.
\bjtitle{Nature}
\bvolume{619}(\bissue{7968}),
\bfpage{52}--\blpage{56}
(\byear{2023})
\doiurl{10.1038/s41586-023-06101-9}
\end{barticle}
\endbibitem

\bibitem[\protect\citeauthoryear{Fukunaga et~al.}{2023}]{PhysRevResearch.5.023054}
\begin{barticle}
\bauthor{\bsnm{Fukunaga}, \binits{R.}},
\bauthor{\bsnm{Haku}, \binits{S.}},
\bauthor{\bsnm{Hayashi}, \binits{H.}},
\bauthor{\bsnm{Ando}, \binits{K.}}:
\batitle{Orbital torque originating from orbital \uppercase{H}all effect in \uppercase{Z}r}.
\bjtitle{Phys. Rev. Res.}
\bvolume{5},
\bfpage{023054}
(\byear{2023})
\doiurl{10.1103/PhysRevResearch.5.023054}
\end{barticle}
\endbibitem

\bibitem[\protect\citeauthoryear{Seifert et~al.}{2023}]{seifert2023time}
\begin{barticle}
\bauthor{\bsnm{Seifert}, \binits{T.S.}},
\bauthor{\bsnm{Go}, \binits{D.}},
\bauthor{\bsnm{Hayashi}, \binits{H.}},
\bauthor{\bsnm{Rouzegar}, \binits{R.}},
\bauthor{\bsnm{Freimuth}, \binits{F.}},
\bauthor{\bsnm{Ando}, \binits{K.}},
\bauthor{\bsnm{Mokrousov}, \binits{Y.}},
\bauthor{\bsnm{Kampfrath}, \binits{T.}}:
\batitle{Time-domain observation of ballistic orbital-angular-momentum currents with giant relaxation length in tungsten}.
\bjtitle{Nat. Nanotechnol.}
\bvolume{18},
\bfpage{1132}--\blpage{1138}
(\byear{2023})
\doiurl{10.1038/s41565-023-01470-8}
\end{barticle}
\endbibitem

\bibitem[\protect\citeauthoryear{Xu et~al.}{2024}]{xu2023orbitronics}
\begin{barticle}
\bauthor{\bsnm{Xu}, \binits{Y.}},
\bauthor{\bsnm{Zhang}, \binits{F.}},
\bauthor{\bsnm{Fert}, \binits{A.}},
\bauthor{\bsnm{Jaffres}, \binits{H.-Y.}},
\bauthor{\bsnm{Liu}, \binits{Y.}},
\bauthor{\bsnm{Xu}, \binits{R.}},
\bauthor{\bsnm{Jiang}, \binits{Y.}},
\bauthor{\bsnm{Cheng}, \binits{H.}},
\bauthor{\bsnm{Zhao}, \binits{W.}}:
\batitle{Orbitronics: \uppercase{L}ight-induced orbit currents in terahertz emission experiments}.
\bjtitle{Nat. Commun.}
\bvolume{15},
\bfpage{2043}
(\byear{2024})
\doiurl{10.1038/s41467-024-46405-6}
\end{barticle}
\endbibitem

\bibitem[\protect\citeauthoryear{An et~al.}{2016}]{an2016spin}
\begin{barticle}
\bauthor{\bsnm{An}, \binits{H.}},
\bauthor{\bsnm{Kageyama}, \binits{Y.}},
\bauthor{\bsnm{Kanno}, \binits{Y.}},
\bauthor{\bsnm{Enishi}, \binits{N.}},
\bauthor{\bsnm{Ando}, \binits{K.}}:
\batitle{Spin--torque generator engineered by natural oxidation of \uppercase{C}u}.
\bjtitle{Nat. Commun.}
\bvolume{7}(\bissue{1}),
\bfpage{13069}
(\byear{2016})
\doiurl{10.1038/ncomms13069}
\end{barticle}
\endbibitem

\bibitem[\protect\citeauthoryear{An et~al.}{2023}]{an2023electrical}
\begin{barticle}
\bauthor{\bsnm{An}, \binits{T.}},
\bauthor{\bsnm{Cui}, \binits{B.}},
\bauthor{\bsnm{Zhang}, \binits{M.}},
\bauthor{\bsnm{Liu}, \binits{F.}},
\bauthor{\bsnm{Cheng}, \binits{S.}},
\bauthor{\bsnm{Zhang}, \binits{K.}},
\bauthor{\bsnm{Ren}, \binits{X.}},
\bauthor{\bsnm{Liu}, \binits{L.}},
\bauthor{\bsnm{Cheng}, \binits{B.}},
\bauthor{\bsnm{Jiang}, \binits{C.}}, \betal:
\batitle{Electrical manipulation of orbital current via oxygen migration in \uppercase{N}i$_{81}$\uppercase{F}e$_{19}$/\uppercase{C}u\uppercase{O}$_x$/\uppercase{T}a\uppercase{N} heterostructure}.
\bjtitle{Adv. Mater.}
\bvolume{35}(\bissue{25}),
\bfpage{2300858}
(\byear{2023})
\doiurl{10.1002/adma.202300858}
\end{barticle}
\endbibitem

\bibitem[\protect\citeauthoryear{Rothschild et~al.}{2022}]{PhysRevB.106.144415}
\begin{barticle}
\bauthor{\bsnm{Rothschild}, \binits{A.}},
\bauthor{\bsnm{Am-Shalom}, \binits{N.}},
\bauthor{\bsnm{Bernstein}, \binits{N.}},
\bauthor{\bsnm{Meron}, \binits{M.}},
\bauthor{\bsnm{David}, \binits{T.}},
\bauthor{\bsnm{Assouline}, \binits{B.}},
\bauthor{\bsnm{Frohlich}, \binits{E.}},
\bauthor{\bsnm{Xiao}, \binits{J.}},
\bauthor{\bsnm{Yan}, \binits{B.}},
\bauthor{\bsnm{Capua}, \binits{A.}}:
\batitle{Generation of spin currents by the orbital \uppercase{H}all effect in \uppercase{C}u and \uppercase{A}l and their measurement by a ferris-wheel ferromagnetic resonance technique at the wafer level}.
\bjtitle{Phys. Rev. B}
\bvolume{106},
\bfpage{144415}
(\byear{2022})
\doiurl{10.1103/PhysRevB.106.144415}
\end{barticle}
\endbibitem

\bibitem[\protect\citeauthoryear{Hayashi et~al.}{2014}]{PhysRevB.89.144425}
\begin{barticle}
\bauthor{\bsnm{Hayashi}, \binits{M.}},
\bauthor{\bsnm{Kim}, \binits{J.}},
\bauthor{\bsnm{Yamanouchi}, \binits{M.}},
\bauthor{\bsnm{Ohno}, \binits{H.}}:
\batitle{Quantitative characterization of the spin-orbit torque using harmonic \uppercase{H}all voltage measurements}.
\bjtitle{Phys. Rev. B}
\bvolume{89},
\bfpage{144425}
(\byear{2014})
\doiurl{10.1103/PhysRevB.89.144425}
\end{barticle}
\endbibitem

\bibitem[\protect\citeauthoryear{Zhu et~al.}{2021}]{PhysRevLett.126.107204}
\begin{barticle}
\bauthor{\bsnm{Zhu}, \binits{L.}},
\bauthor{\bsnm{Zhu}, \binits{L.}},
\bauthor{\bsnm{Buhrman}, \binits{R.A.}}:
\batitle{Fully spin-transparent magnetic interfaces enabled by the insertion of a thin paramagnetic nio layer}.
\bjtitle{Phys. Rev. Lett.}
\bvolume{126},
\bfpage{107204}
(\byear{2021})
\doiurl{10.1103/PhysRevLett.126.107204}
\end{barticle}
\endbibitem

\bibitem[\protect\citeauthoryear{Zhu et~al.}{2019}]{PhysRevApplied.11.061004}
\begin{barticle}
\bauthor{\bsnm{Zhu}, \binits{L.}},
\bauthor{\bsnm{Zhu}, \binits{L.}},
\bauthor{\bsnm{Shi}, \binits{S.}},
\bauthor{\bsnm{Sui}, \binits{M.}},
\bauthor{\bsnm{Ralph}, \binits{D.C.}},
\bauthor{\bsnm{Buhrman}, \binits{R.A.}}:
\batitle{Enhancing spin-orbit torque by strong interfacial scattering from ultrathin insertion layers}.
\bjtitle{Phys. Rev. Appl.}
\bvolume{11},
\bfpage{061004}
(\byear{2019})
\doiurl{10.1103/PhysRevApplied.11.061004}
\end{barticle}
\endbibitem

\bibitem[\protect\citeauthoryear{Nguyen et~al.}{2016}]{nguyen2016enhanced}
\begin{botherref}
\oauthor{\bsnm{Nguyen}, \binits{M.-H.}},
\oauthor{\bsnm{Zhao}, \binits{M.}},
\oauthor{\bsnm{Ralph}, \binits{D.}},
\oauthor{\bsnm{Buhrman}, \binits{R.}}:
Enhanced spin \uppercase{H}all torque efficiency in \uppercase{P}t$_{100-x}$\uppercase{A}l$_x$ and \uppercase{P}t$_{100-x}$\uppercase{H}f$_x$ alloys arising from the intrinsic spin \uppercase{H}all effect.
Appl. Phys. Lett.
\textbf{108}(24)
(2016)
\doiurl{10.1063/1.4953768}
\end{botherref}
\endbibitem

\bibitem[\protect\citeauthoryear{Zhu et~al.}{2018}]{PhysRevApplied.10.031001}
\begin{barticle}
\bauthor{\bsnm{Zhu}, \binits{L.}},
\bauthor{\bsnm{Ralph}, \binits{D.C.}},
\bauthor{\bsnm{Buhrman}, \binits{R.A.}}:
\batitle{Highly efficient spin-current generation by the spin \uppercase{H}all effect in \uppercase{A}u$_{1-x}$\uppercase{P}t$_x$}.
\bjtitle{Phys. Rev. Appl.}
\bvolume{10},
\bfpage{031001}
(\byear{2018})
\doiurl{10.1103/PhysRevApplied.10.031001}
\end{barticle}
\endbibitem

\bibitem[\protect\citeauthoryear{Shi et~al.}{2018}]{PhysRevApplied.9.011002}
\begin{barticle}
\bauthor{\bsnm{Shi}, \binits{S.}},
\bauthor{\bsnm{Ou}, \binits{Y.}},
\bauthor{\bsnm{Aradhya}, \binits{S.V.}},
\bauthor{\bsnm{Ralph}, \binits{D.C.}},
\bauthor{\bsnm{Buhrman}, \binits{R.A.}}:
\batitle{Fast low-current spin-orbit-torque switching of magnetic tunnel junctions through atomic modifications of the free-layer interfaces}.
\bjtitle{Phys. Rev. Appl.}
\bvolume{9},
\bfpage{011002}
(\byear{2018})
\doiurl{10.1103/PhysRevApplied.9.011002}
\end{barticle}
\endbibitem

\bibitem[\protect\citeauthoryear{Zhang et~al.}{2016}]{zhang2016critical}
\begin{botherref}
\oauthor{\bsnm{Zhang}, \binits{C.}},
\oauthor{\bsnm{Fukami}, \binits{S.}},
\oauthor{\bsnm{Watanabe}, \binits{K.}},
\oauthor{\bsnm{Ohkawara}, \binits{A.}},
\oauthor{\bsnm{DuttaGupta}, \binits{S.}},
\oauthor{\bsnm{Sato}, \binits{H.}},
\oauthor{\bsnm{Matsukura}, \binits{F.}},
\oauthor{\bsnm{Ohno}, \binits{H.}}:
Critical role of \uppercase{W} deposition condition on spin-orbit torque induced magnetization switching in nanoscale \uppercase{W}/\uppercase{C}o\uppercase{F}e\uppercase{B}/\uppercase{M}g\uppercase{O}.
Appl. Phys. Lett.
\textbf{109}(19)
(2016)
\doiurl{10.1063/1.4967475}
\end{botherref}
\endbibitem

\bibitem[\protect\citeauthoryear{Ito et~al.}{2022}]{ito2022perpendicularly}
\begin{barticle}
\bauthor{\bsnm{Ito}, \binits{K.}},
\bauthor{\bsnm{Kikuchi}, \binits{N.}},
\bauthor{\bsnm{Seki}, \binits{T.}},
\bauthor{\bsnm{Takanashi}, \binits{K.}}:
\batitle{Perpendicularly magnetized epitaxial \uppercase{C}o/\uppercase{N}i multilayers grown on \uppercase{R}u (0001) layers by alternate monoatomic layer deposition}.
\bjtitle{J. Magn. Magn. Mater.}
\bvolume{563},
\bfpage{169908}
(\byear{2022})
\doiurl{10.1016/j.jmmm.2022.169908}
\end{barticle}
\endbibitem

\bibitem[\protect\citeauthoryear{Oliveira et~al.}{2020}]{oliveira2020simple}
\begin{barticle}
\bauthor{\bsnm{Oliveira}, \binits{F.}},
\bauthor{\bsnm{Cipriano}, \binits{R.}},
\bauthor{\bsnm{Silva}, \binits{F.}},
\bauthor{\bsnm{Rom{\~a}o}, \binits{E.}},
\bauthor{\bsnm{Dos~Santos}, \binits{C.}}:
\batitle{Simple analytical method for determining electrical resistivity and sheet resistance using the van der \uppercase{P}auw procedure}.
\bjtitle{Sci. Rep.}
\bvolume{10}(\bissue{1}),
\bfpage{16379}
(\byear{2020})
\doiurl{10.1038/s41598-020-72097-1}
\end{barticle}
\endbibitem

\bibitem[\protect\citeauthoryear{Lee et~al.}{2013}]{lee2013threshold}
\begin{botherref}
\oauthor{\bsnm{Lee}, \binits{K.-S.}},
\oauthor{\bsnm{Lee}, \binits{S.-W.}},
\oauthor{\bsnm{Min}, \binits{B.-C.}},
\oauthor{\bsnm{Lee}, \binits{K.-J.}}:
Threshold current for switching of a perpendicular magnetic layer induced by spin \uppercase{H}all effect.
Appl. Phys. Lett.
\textbf{102}(11)
(2013)
\doiurl{10.1063/1.4798288}
\end{botherref}
\endbibitem

\end{thebibliography}

\end{document}